# Near-atomic investigation on the elemental redistribution during co-precipitation of nano-sized kappa phase and B2 phase in an Al-alloyed lightweight steel

*Bowen Zou* [a], *Yixu Wang* [a,b], *Xiao Shen* [a], *Philipp Krooß* [c], *Thomas Niendorf* [c], *Richard Dronskowski* [b], *Wenwen Song* [a*]

[a]: Department of Granularity of Structural Information in Materials Engineering, Institute of Materials Engineering, University of Kassel, Mönchebergstr. 3, 34125 Kassel, Germany

[b]: Chair of Solid-State and Quantum Chemistry, Institute of Inorganic Chemistry, RWTH Aachen University, Landoltweg 1, 52056 Aachen, Germany

[c]: Department of Metallic Materials, Institute of Materials Engineering, University of Kassel, Mönchebergstr. 3, 34125 Kassel, Germany

[*] Corresponding author: song@uni-kassel.de

**Abstract:**

In the present study, correlative transmission Kikuchi diffraction-transmission electron microscopy (TKD-TEM) measurements, atom probe tomography (APT) and density functional theory (DFT) calculations are used to reveal the elemental redistribution during co-precipitation of nano-sized kappa ($\kappa$) and B2 phases in an FCC matrix in an Al-alloyed Fe-10Al-7Mn-6Ni-1C (wt.%) steel. Upon ageing at 800 °C for 15 min, two co-nanoprecipitation modes are observed: B2 forming together with $\kappa$ and B2 forming separately from $\kappa$ in the FCC matrix. APT reveals that the B2 precipitate next to $\kappa$ (referred to as $B2_I$) is close to an FeAl-type phase, while the isolated B2 precipitate (referred to as $B2_{II}$) is close to a NiAl-type phase. The $\kappa$ precipitates maintain a nearly constant Al content (≈18.4 at.%) regardless of their precipitation position. DFT confirms that $\kappa$ may accommodate limited Ni substitution at Fe sites without losing structural stability, and that Fe–Ni atoms exchange between $\kappa$ and B2 is thermodynamically favorable at 800 °C. This exchange drives the B2 phase to evolve from NiAl-type towards FeAl-type, improving the stability of both phases during co-precipitation. These results provide understanding of $\kappa$ - B2 interactions and offer insights for designing nanosized intermetallic-strengthened microstructures in Al-alloyed lightweight steels.

## 1. Introduction

Advanced high-strength steels have been used in the automotive industry for decades, thanks to their variety of strength levels and ductility that meet most requirements for automotive structures [1,2]. Assuming that the transitions from fuel-powered vehicles to electric ones, the use of heavy batteries makes it challenging to reduce the entire vehicle's weight without affecting safety or raising costs. Aluminum (Al)-alloyed steels provide a good solution for manufacturing lightweight components for the automotive sector due to their low density, cost-effectiveness, and superior mechanical properties compared to lighter alloys such as aluminum and magnesium alloys [1,3,4]. For decades, studies [5–11] have focused on Al-alloyed intermetallic-strengthened steels for lightweight applications. These steels contain 10-28 wt.% Mn, 5-12 wt.% Al, and 0.7-1.2 wt.% C, which exhibits a microstructure composed of face-centered cubic (FCC) or FCC plus body-centered cubic (BCC) matrices. The addition of more than 8.5 wt.% Al can reduce steel density by around 10% [12]. More importantly, Al enables stabilization of intermetallic phases, notably the kappa ($\kappa$) and B2 phases, in these Al-alloyed lightweight steels. It has been reported that uniformly distributed nano-sized $\kappa$ precipitates in the FCC matrix enhance strength and hardness without sacrificing ductility by promoting shear-band induced plasticity (SIP) and glide plane softening through dislocation glide along {111} planes [13–15]. Meanwhile, the presence of B2 phases within the BCC matrix increases strength despite a risk of embrittlement, because dislocations shear these coherent B2 precipitates, forming anti-phase boundaries that impede dislocation motion [16–18].

Earlier studies primarily concentrated on the phase-specific precipitation of $\kappa$ or B2 phases. In contrast, recent research [19–24] has shifted the attention to the co-precipitation of $\kappa$ and B2 phases within the FCC matrix of lightweight steels, aiming to improve both strength and ductility simultaneously. The primary motivation for introducing co-nanoprecipitation in the FCC matrix is to realize the complementary strengthening characteristics of these precipitates: $\kappa$-precipitates, when small and coherent, are shearable and can enhance strength by promoting SIP; whereas incoherent B2 precipitates, being non-shearable, impede dislocation movement through Orowan mechanisms, thereby improving strain-hardening capabilities and preventing strain localization. With increasing size, $\kappa$-precipitates may gradually lose coherency and contribute to strengthening also via Orowan looping. Several studies clearly illustrate this concept. Kies et al. [19] and Moon et al. [20] demonstrated that Ni addition promotes the co-precipitation of $\kappa$ and B2 phases in Al-alloyed Fe-Al-Mn-Ni-C lightweight steels. In addition, several studies have explored co-nanoprecipitation strategies in Al-alloyed lightweight steels, reporting that a homogeneous dispersion of $\kappa$ and B2 precipitates significantly improves the strength–ductility balance and strain-hardening capability. For example, An et al. [21] and Wang et al. [22] highlighted the synergistic roles of planar slip and Orowan strengthening, while Zhang et al. [23] and Burja et al. [24] emphasized the importance of refined and uniformly distributed $\kappa$ along with precisely controlled nanosized B2 phases. Together, these studies conclude that co-nanoprecipitation of $\kappa$ and B2 phases is an effective approach to improve strength and ductility in Al-alloyed lightweight steels simultaneously.

Despite extensive investigation on the mechanical enhancement mechanisms attributed to co-nanoprecipitation of $\kappa$ and B2 phases, however, fundamental understanding of the interaction between them during co-nanoprecipitation remains limited. Zargaran et al. [25] specifically addressed $\kappa$-assisted nucleation of B2, demonstrating that pre-existing $\kappa$-nanoprecipitates act as nucleation sites, effectively promoting uniform distribution of nano-sized B2 within the FCC matrix. This strategy successfully improves both strength and ductility by optimizing the morphology and distribution of B2 precipitates. Nevertheless, this study mainly focused on the nucleation assistance of $\kappa$ phase for B2 but did not explore the elemental partitioning behavior or potential competition for solute elements between $\kappa$ and B2 phases during co-precipitation. Major questions remain unclear, such as how the different co-precipitation modes of $\kappa$ and B2 phases in the FCC matrix, e.g., either $\kappa$ and B2 precipitates forming separately or growing next to each other, affect their morphology and local chemical composition. In other words, how does elemental redistribution, particularly of Ni and Al, occur when $\kappa$ and B2 precipitates are spatially close and interact with each other? Do these intermetallic phases compete for solute elements or assist in the stabilization of each other? Addressing these knowledge gaps is essential for further optimizing co-nanoprecipitation strategies and fully digging into their potential to enhance mechanical properties in Al-alloyed Fe-Mn-Al-Ni-C lightweight steels.

In the present work, we combined correlative transmission Kikuchi diffraction-transmission electron microscopy (TKD-TEM), atom probe tomography (APT) and density functional theory (DFT) calculations to investigate the nano characteristics of $\kappa$ and B2 phases and the elemental redistribution between them when they co-nanoprecipitate in an FCC matrix. Specifically, we revealed the co-precipitation of nanosized $\kappa$ and B2 phases in the FCC matrix near the FCC/BCC phase boundary. Then, we studied the interaction between $\kappa$ and B2 phases during their co-nanoprecipitation, focusing on the elemental partitioning of Fe, Ni, and Al using APT. Additionally, DFT-based calculations were employed to evaluate the tolerance of $\kappa$ phases to elemental substitution (such as Fe/Ni/Al site exchange), providing insights into their thermodynamic stability and clarifying the precipitation behavior influenced by interactions between intermetallic phases.

## 2. Materials and methods

### 2.1. Materials and processing

The investigated Al-alloyed lightweight steel, with a nominal chemical composition of Fe-10Al-7Mn-6Ni-1C (wt.%), exhibits a duplex matrix microstructure consisting of FCC and BCC phases. The material underwent vacuum-induction melting into ingots, followed by homogenization at 1200 °C for 5 hours, and then multi-step open-die hot forging into slabs before undergoing an annealing treatment. A detailed description of alloy design and materials fabrication has been reported in previous studies [26,27]. The exact chemical composition of the investigated steel, including the major alloying elements and impurities, was determined by wet-chemical analysis and is provided in **Table 1** in weight percent (wt.%) and atomic percent (at.%). To achieve full dissolution of intermetallic phases after hot forging, a solution annealing treatment at 1200 °C for 2 hours was conducted. Subsequent water quenching was intended to restrain the reprecipitation of intermetallic phases during cooling. Subsequent isothermal ageing was performed at 800 °C for 15 minutes, followed by water quenching to room temperature [27].

In our previous work [27], the microstructure evolution of this steel during isothermal ageing at 800 °C was characterized using Scanning Electron Microscopy (SEM), Electron Backscatter Diffraction (EBSD), Energy Dispersive X-ray Spectroscopy (EDS), and mechanical testing. In the solution-annealed state before ageing, SEM and EBSD revealed a microstructure consisting of ≈87% FCC and ≈13% BCC phase. EDS indicated Al and Ni enrichment and Mn depletion in the BCC matrix. Mechanical testing revealed that after 15 minutes of ageing, an excellent combination of yield strength ($\sigma_{YS} \approx$ 1300 MPa) and elongation at fracture ($\varepsilon_f \approx$ 50) was achieved. In contrast, prolonged ageing to 60 minutes led to a deterioration of ductility due to extensive precipitation and possible over-ageing. In the present work, APT and DFT investigations are conducted to study the local chemical composition and elemental redistribution of $\kappa$ and B2 phases at 800 °C for 15 minutes, to elucidate their co-nanoprecipitation behavior.

**Table 1.** Chemical composition (in wt.% & at.%) of the investigated Al-alloyed steel.

| Al-alloyed Steel | Fe | Al | Ni | Mn | C | Si | P | S | N |
|---|---|---|---|---|---|---|---|---|---|
| wt.% | balance | 10.10 | 6.10 | 6.80 | 0.95 | 0.025 | 0.012 | 0.011 | 0.012 |
| at.% | balance | 18.31 | 5.08 | 6.06 | 3.87 | 0.040 | 0.020 | 0.020 | 0.040 |

### 2.2. Methods

From the FCC/BCC interface, a site-specific FIB lift-out procedure was adopted using a FEI Helios Nanolab 660 FIB/SEM dual-beam system to prepare a TEM lamella. After reducing the thickness of the TEM lamella, the TEM specimen was fixed on a pretilt holder for TKD measurement in order to correlate the phase identification of matrixes with the TEM images. During TKD measurement in the FIB/SEM dual-beam system, the pre-tilt sample holder together with the working stage was tilted 15° to enable the signal to be captured by the EBSD detector. TKD analysis was conducted by means of the software texture & elemental analytical microscopy (TEAM) with a step size of 50 nm. Further data analysis was conducted using the software OIM analysis V8. Subsequently, TEM measurements of nanosized $\kappa$ and B2 phases in the FCC matrix were carried out using a FEI Tecnai 20 at 300 kV.

APT was used to investigate the characteristics and local chemical composition of $\kappa$ and B2 precipitates in depth, as well as the elemental partitioning behavior across different interfaces. The needle shape sample preparation for APT measurement was conducted using the FEI Helios Nanolab 660 dual-beam FIB-SEM system mentioned before. The three-dimensional (3D) characterization and local chemical composition analysis of the matrices and intermetallic $\kappa$ and B2 phases were performed by a high-resolution Local Electrode Atom Probe (LEAP) 4000X HR system (CAMECA Instrument Inc.) with a reflectron lens for improved mass resolution. The APT specimens were measured in the laser-pulse mode with the evaporation rate of 0.5%. The laser pulse energy was 30 pJ, and the pulse frequency was set to 250 kHz. In the analysis chamber, the base temperature was kept at 60 K, and the pressure was controlled below $6 \times 10^{-11}$ torr. The 3D reconstruction of the collected APT data and further analysis of the chemical composition were conducted using software AP suite 6.1 from CAMECA.

The density-functional calculations were performed using the Vienna ab initio simulation package (VASP, version: 6.4.1) [28,29]. Projector augmented wave (PAW) pseudopotentials [30] with generalized gradient approximation (GGA) exchange-correlation (XC) functionals with the Perdew-Burke-Ernzerhof (PBE) parameterizations [31] were used. The plane-wave kinetic energy cutoff was set to 800 eV. Spin polarization was considered throughout the work. Initial magnetic moments for transition-metal atoms were set to be 5 $\mu_B$. For structural optimizations, Gaussian smearing with a smearing width of 0.01 eV was employed to account for the fractional occupancy of electrons in the vicinity of the Fermi level ($E_F$). Convergence of optimization was reached upon the difference of energies and forces of consecutive iterations fell below $10^{-8}$ eV and $10^{-6}$ eV/Å, respectively. The same energy convergence criterion was kept for the subsequent calculations. A Γ-centered Monkhorst-Pack $k$ point mesh with a mesh grid of 0.02× 2π Å$^{-1}$ was automatically generated. As the structures were sufficiently optimized, the pressure ($p$) was set to be 0. At ground state, the internal energy ($U$) and enthalpy (defined by $H = U + pV$, where $V$ is the volume of the system) of a given system were represented by its total energy ($E$). Regarding the thermodynamic calculations at finite temperature, they were conducted within the harmonic oscillator approximations and implemented through a finite displacement scenario. Configurations to calculate force constants were generated by phonopy [32] and the DFT results from VASP were harvested by the same code to calculate the partition functions [33]. Based on the plane-wave results, a reciprocal space unitary transformation was performed by Local-Orbital Basis Suite Towards Electronic-Structure Reconstruction [34–36] (LOBSTER). Subsequently, crystal orbital Hamilton population (COHP) analysis [37] was performed to elucidate the interactions between atomic pairs in terms of their energetic contributions. The results from LOBSTER were subsequently processed by the post-processing software LOPOSTER [38].

## 3. Results and discussion

### 3.1 Characterization of intermetallic $\kappa$ and B2 phases using correlative TKD-TEM measurements

The nanoscale characteristics of the $\kappa$ and B2 phases in the investigated Al-alloyed lightweight steel after ageing at 800 °C for 15 minutes were initially investigated using correlative TKD–TEM measurements. The TKD analysis was performed directly after cross-sectional TEM specimen preparation using FIB, facilitating phase identification of the matrices in the TEM lamella. The TKD phase map shown in **Figure 1**a reveals a microstructure consisting of FCC (red) and BCC (green) phases, supporting the phase identification of the matrices during TEM measurements. However, the limited spatial resolution resulting from the 50 nm step size in the TKD measurement restricted direct characterization of both intermetallic phases. The small dots seen within the FCC and BCC matrices were noise signals generated during TKD data acquisition.

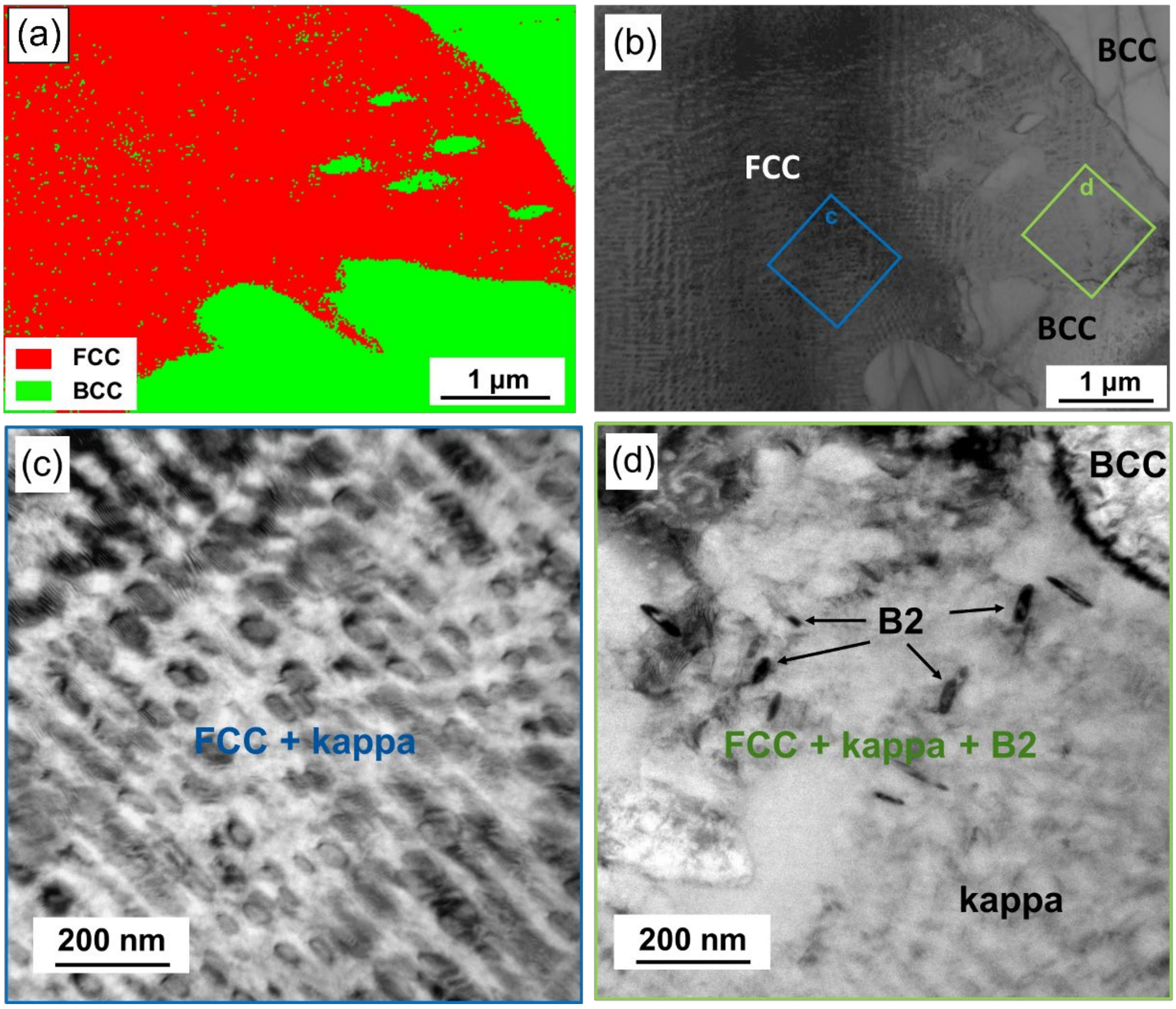

**Figure 1.** TKD phase map and TEM bright-field image of the investigated steel after 15 min ageing at 800 °C. (a) Correlative phase identification of the TEM specimen via TKD, the red area indicates the FCC matrix, and the green area indicates the BCC matrix; (b) Correlative bright-field image of FCC and BCC matrices; (c) and (d) TEM bright-field image of *κ* and B2 phases in the FCC matrix.

In correspondence with the TKD phase map, the bright-field TEM image of the duplex matrices and the secondary phases in the FCC matrix after ageing at 800 °C for 15 minutes is shown in Figure 1b. As illustrated in the magnified regions in Figure 1c and 1d, nano-sized cuboidal precipitates are homogeneously distributed within the FCC matrix, separated by narrow channels. After solution annealing or ageing, a similar morphology of cuboidal precipitates in the FCC matrix was observed in numerous studies and identified as *κ* phase via selected area electron diffraction of TEM measurements [39,40]. Therefore, these cuboidal precipitates are presumed to be nano-sized *κ* phases in our study. The cuboidal *κ* precipitates are aligned into particle stacks, which ultimately form a network-like arrangement throughout the entire FCC matrix. In the present work, the cross-section of the *κ* precipitates perpendicular to the stacking direction is approximately 50 to 90 nm in length, while their thickness parallel to the stacking direction is on the order of 15 to 50 nm. Moreover, Figure 1d reveals that, in addition to the cuboidal *κ* precipitates, co-precipitated nano-sized disc-like precipitates are present within the same region of the FCC matrix and near the FCC/BCC phase boundary. These disc-like precipitates have lengths ranging from 48 to 87 nm and thicknesses of approximately 12 to 22 nm. They are presumably B2 precipitates and has similar morphology and size as the incoherent B2 precipitates shown in the FCC matrix in An`s studies [41,42] of Fe–28Mn–11Al–1C–5Ni (wt.%) steel. To obtain a more detailed understanding of the morphology and local chemical composition of these precipitates within the FCC matrix, APT analyses were conducted.

### 3.2 Near-atomic study of the co-precipitation of *κ* and B2 phases in the FCC matrix in the studied Al-alloyed lightweight steel by APT

Near-atomic scale APT analysis was conducted to characterize the co-precipitation of the assumed *κ* and B2 phases in the FCC matrix of the investigated steel Fe-10Al-7Mn-6Ni-1C (wt.%) after ageing at 800 °C for 15 min. In **Figure 2**, two APT tips are compared to study the intermetallic precipitates in different regions of the FCC matrix. Figure 2a shows the APT reconstruction of the specimen lifted out from the interior of the FCC matrix, which includes the 3D reconstruction of the cuboidal *κ* precipitates embedded in the FCC matrix. The elemental maps demonstrate the localized enrichment of C and Al, indicating *κ* precipitates in the FCC matrix. The distinction between the Al enrichment and the Al depletion areas in the Al elemental mapping is not obvious. In contrast, the C-deficient and C-rich phases in C elemental mapping are attributed to the FCC matrix and *κ* precipitates, respectively. The Mn and Ni yield a homogeneous distribution in the FCC matrix. The reconstructed *κ*/FCC interfaces are highlighted in green using iso-concentration surfaces that enclose regions with a C concentration above 6 at.%. The cuboidal *κ* precipitates are densely packed into particle stacks along orthogonal directions. To further analyze each *κ* precipitate, the *κ* precipitates in this APT tip were labelled as $K_1$ to $K_7$. In order to evaluate the elemental partitioning behavior between *κ* precipitates and the FCC matrix, regions of interest (ROI) analyses were performed across *κ* precipitates and the channels between them. We present a representative 1D concentration profile obtained from ROI 1, which traverses through precipitates $K_5$ to $K_7$. Compared to the FCC channel, the strong Al- and C-enrichment, as well as the depletion of Fe, indicate the presence of *κ* precipitates. Simultaneously, the Mn and Ni exhibit a slight depletion in *κ* precipitates. The *κ* phase in the FCC matrix showed an average chemical composition of 61.2 at.% Fe, 18.4 at.% Al, 6.2 at.% Mn, 4.6 at.% Ni, 9.2 at.% C, roughly corresponding to an approximate stoichiometry of $(Fe_{2.7},Mn_{0.27},Ni_{0.2})Al_{0.8}C_{0.40}$ in a L'$1_2$-structure, while the adjacent FCC channels exhibit an average composition of 68.9 at.% Fe, 15.6 at.% Al, 6.4 at.% Mn, 5.3 at.% Ni, 3.3 at.% C. In comparison to the ideal stoichiometry of *κ* phase $(Fe,Mn)_{3.0}Al_{1.0}C_{1.0}$ [43,44], the slight Al deficiency and the strong C off-stoichiometry indicate the existence of a certain amount of Al vacancies and a high number of C vacancies in the *κ* phase. Besides, this stoichiometry pinpoints a partial substitution of Fe-sites by Mn and Ni, as well as a minor substitution of Al-sites by Mn.

Figure 2b displays the analysis of the APT specimen taken at the FCC/BCC phase boundary, including a 3D atom map highlighting *κ* and B2 precipitates near the phase boundary. As shown in the C and Mn atom maps, there is significant C and Mn partitioning across the full tip. Both the atom maps and the 1D concentration profile of selected ROI 2 demonstrate that the local C and Mn enrichment is at the lower left-hand region, while the upper right-hand region is strongly depleted in these elements. Thus, the full APT tip consists of two important regions, namely the intermetallic phases within FCC region located near the FCC/BCC phase boundary and the BCC matrix region. The FCC matrix exhibits an average composition of 69.1 at.% Fe, 13.6 at.% Al, 9.3 at.% Mn, 3.2 at.% Ni, and 3.2 at.% C. In contrast, the BCC matrix shows a composition of 71.9 at.% Fe, 17.5 at.% Al, 5.4 at.% Mn, 4.5 at.% Ni, and a significantly lower C content of 0.04 at.%. The Mn content is lower in the interior of the FCC matrix compared to the region near the phase boundary, while the contents of Al, Ni, and C are slightly higher inside the matrix. This may be attributed to Mn heterogeneity at the phase boundary during its partitioning into the FCC matrix [45], whereas C, Al and Ni, which tend to stabilize intermetallic phases [46–48], contribute to *κ* and B2 co-precipitation in the FCC matrix close to the phase boundary. Consistent with this, local enrichment of C and Ni in the elemental maps was observed near the phase boundary, pointing to the presence of *κ* and B2 co-precipitation within this region. In the 3D reconstruction, these presumed intermetallic phases are visualized using iso-concentration surfaces with 6 at.% C (green) and 12 at.% Ni (blue), respectively.

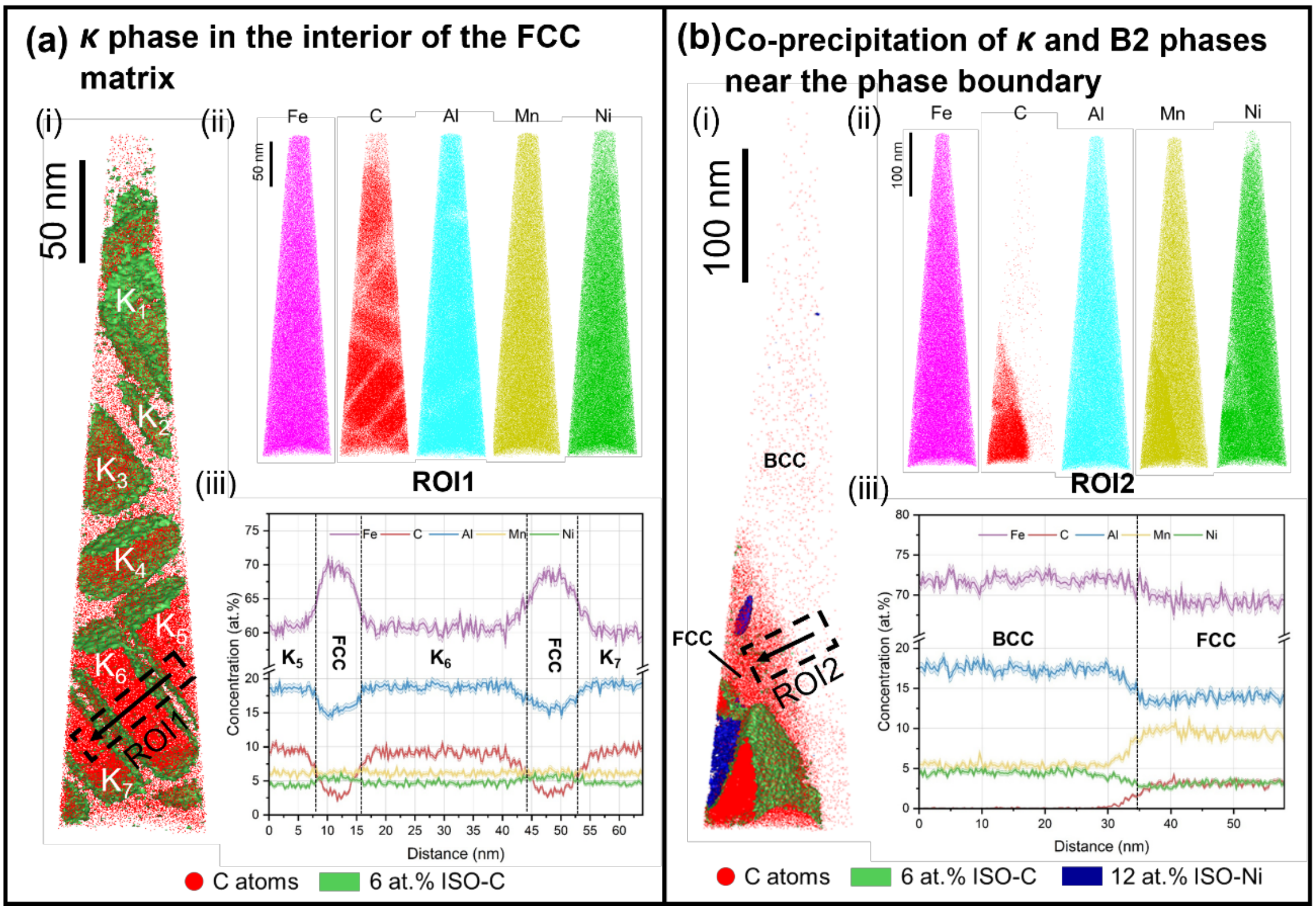

**Figure 2.** (a) APT tip lifted out from the interior of the FCC matrix: i) 3D reconstruction of cuboidal $\kappa$ precipitates embedded in the FCC matrix; ii) elemental mapping of Fe, C, Al, Mn, Ni; iii) 1D concentration profile of the selected ROI through $\kappa$ precipitates and channels among them. (b) APT tip extracted at the FCC/BCC phase boundary: i) 3D reconstruction of $\kappa$ and B2 precipitates near the phase boundary; ii) elemental mapping of Fe, C, Al, Mn, Ni; iii) 1D concentration profile of the selected ROI through the FCC/BCC phase boundary.

To further investigate the characteristics of $\kappa$ and B2 precipitates within the FCC matrix near the phase boundary, additional analyses were conducted from different views and selected ROIs, focusing on two co-precipitation modes, namely B2 forming together with $\kappa$ or B2 forming separately from $\kappa$ in the FCC matrix. **Figure 3** shows the 3D reconstruction of $\kappa$ and B2 phases in the full C atom map (Figure 3a) and the magnified view of intermetallic phases/FCC region derived from three perspectives (Figure 3b). In this region, three C-rich precipitates are identified and referred to as $K_1$-$K_3$, while two Ni-rich precipitates are named as $B2_I$ and $B2_{II}$ for distinction. It is worth noting that $B2_I$ resides in the channel between $K_1$ and $K_2$, exhibiting coordinated growth and direct contact with $\kappa$ precipitates. In contrast, the $B2_{II}$ precipitated within the FCC matrix but separately from $\kappa$ precipitates. Although the Ni-rich precipitates are only partially visualized in this APT tip, the length of the $B2_I$ exceeds 200 nm, with a thickness well above 20 nm, while the $B2_{II}$ shows a thickness of around 10 nm. Furthermore, the C-rich $\kappa$ precipitates exhibit a polyhedral morphology with a near-cuboidal shape. Based on the relatively complete shapes of $K_2$ and $K_3$, as shown in Figure 3b, their dimensions are estimated to be approximately 100 to 110 nm in edge length and 40 to 50 nm in thickness.

Three selected cylindrical ROIs are used to assess the atomic-scale chemistry of the B2 precipitates, the FCC matrix, and $\kappa$ precipitates, as well as the elemental partitioning behavior between $B2_I$-$K_2$, $B2_{II}$-FCC, and between $K_2$-$K_3$. The corresponding 1D concentration profiles for each ROI are shown in Figure 3c-e. The 1D concentration profile of the Ni- and Al-enriched precipitate associated with ROI 3 (Figure 3c) exhibits a gradient in chemical composition at the $B2_I$-$K_2$ interface. The $B2_I$ yields a strong Al- and Ni-enrichment and depletion of Mn and C. The average chemistry of $B2_I$ is 51.5 at.% Fe, 24.8 at.% Al, 4.0 at.% Mn, 14.5 at.% Ni and 0.3 at.%

C, with an approximate stoichiometry of $(Fe_{0.54}Ni_{0.15}Mn_{0.04})Al_{0.26}$. According to the phase diagram of FeAl [49], the FeAl-type B2 phase at 800 °C can tolerate significant off-stoichiometry, with Fe:Al ratios reaching up to 75:25. This composition range is commonly referred to as a defected B2 structure (B2`). The presence of 14.5 at.% Ni and 4.0 at.% Mn in $B2_I$ indicates partial substitution of Fe by Ni and Mn. Therefore, the observed chemical stoichiometry can be interpreted as a variant of FeAl with a defected B2 structure at 800 °C, which deviates from the ideal stoichiometry and reflects the complex substitution behavior in this system. Figure 3d displays the 1D concentration profile of ROI4 across $B2_{II}$-FCC, exhibiting a strong Ni- and Al-enrichment with an average chemical concentration of 37.3 at.% and 34.8 at.%, respectively. Besides, the concentration of Fe, Mn and C decreases to 22.2 at. %, 3.7 at.% and 0.7 at.%, respectively. The near 50:50 distribution of Ni and Al indicates that $B2_{II}$ corresponds compositionally to a NiAl-type phase, with an average stoichiometry of $Ni_{0.38}Al_{0.36}Fe_{0.23}Mn_{0.04}$. Although the 1D concentration profile cannot by itself confirm the crystal structure, its Ni, Al-rich and Fe, Mn, C-depleted composition supports the interpretation of $B2_{II}$ as a B2-ordered phase, which is consistent with APT-resolved B2 precipitates showing a near-equiatomic Ni:Al ratio and negligible C in related Fe-Mn-Al-Ni-C steels [19,22]. Evidence for the B2-ordered phase in such systems has been further provided by superlattice reflections in selected area electron diffraction and synchrotron X-ray diffraction [19]. Moreover, the observed deviation from ideal 1:1 Ni:Al stoichiometry in the B2 phase, particularly the deficiency in Al and excess in Ni or Fe, points at Mn as a substitution of Al and the possible presence of Al vacancies. Such non-stoichiometric NiAl-B2 phases ($Ni_{1+x}Al_{1-x}$) containing Al-site vacancies have been frequently reported and are thermodynamically stable under certain thermal and compositional conditions in Fe-Mn-Al-Ni-C steels [50,51]. Based on the 1D concentration profile of ROI 5 through $K_2$ – $K_3$ (Figure 3e), the average atomic composition of Fe, Al, Mn, Ni, C in the $\kappa$ phase was determined as 61.2 at.%, 18.4 at.%, 6.2 at.%, 4.6 at.%, 9.2 at.%. Assuming that the face-centered sublattice (occupied by Fe, Mn, or Ni) contains three atoms per unit, the approximate stoichiometry of the $\kappa$ phase near the FCC/BCC phase boundary can be expressed as $(Fe_{2.5}Mn_{0.3}Ni_{0.19})Al_{0.8}C_{0.42}$. This representation indicates a minor substitution of Fe sites by Ni and Mn, along with the presence of both Al and C vacancies. Compared to Al, the C off-stoichiometry is more pronounced with an estimated occupation of 0.42. A summary of the atomic compositions of the $\kappa$ and B2 precipitates determined by APT in this investigated steel is provided in supplementary **Tables S1** and **S2**.

The $\kappa$ phase precipitated near the FCC/BCC phase boundary shows a slightly higher Mn and C content compared to that observed within the FCC matrix interior. This enrichment is likely related to the partitioning of Mn and especially C into FCC and their accumulation close to the interface, where both elements are known to stabilize the $\kappa$ phase. More interestingly, two types of B2 precipitates were identified in the FCC matrix near the phase boundary, exhibiting position-dependent chemical characteristics. The $B2_I$, which forms together with neighboring $\kappa$ precipitates ($K_1$ and $K_2$), is identified as an FeAl-type B2 phase with relatively lower Al and Ni contents than those in $B2_{II}$. In contrast, the $B2_{II}$ forms separately from $\kappa$ precipitates within the FCC matrix and appears as a Fe-containing NiAl-type phase with a Ni:Al compositional ratio near 1:1. The 1D profile of ROI 3 (Figure 3c) reveals that in the $B2_I$–$K_2$ interface, significant solute disproportion occurs, particularly involving Fe, Al, and Ni, indicating a local element exchange during co-precipitation and coordinated growth of $B2_I$ and $K_2$. ROI 4 (Figure 3d), in comparison, shows elemental partitioning at the $B2_{II}$–FCC interface, where the $\kappa$ phase is not involved. These findings imply that B2 precipitates within the FCC matrix likely have different formation mechanisms according to their positional relationship with the $\kappa$ phase. The presence or absence of adjacent $\kappa$ precipitates appears to strongly influence the chemical composition of the B2 phase, potentially modifying their precipitation behaviors. However, it remains unclear how elemental redistribution occurs between B2 phases and the adjacent $\kappa$ phase, and whether they compete with or assist each other in different precipitation scenarios.

To address this, further investigation into atomic-scale mechanisms governing elemental partitioning behaviors between $\kappa$ and B2 phases and their thermodynamic stability is required. In the following section, DFT-based calculations are considered to evaluate the tolerance of $\kappa$ and B2 phases to elemental substitution (e.g., Fe-Ni atom exchange), providing insights into their thermodynamic stability and an atomic-level understanding of their co-precipitation behavior.

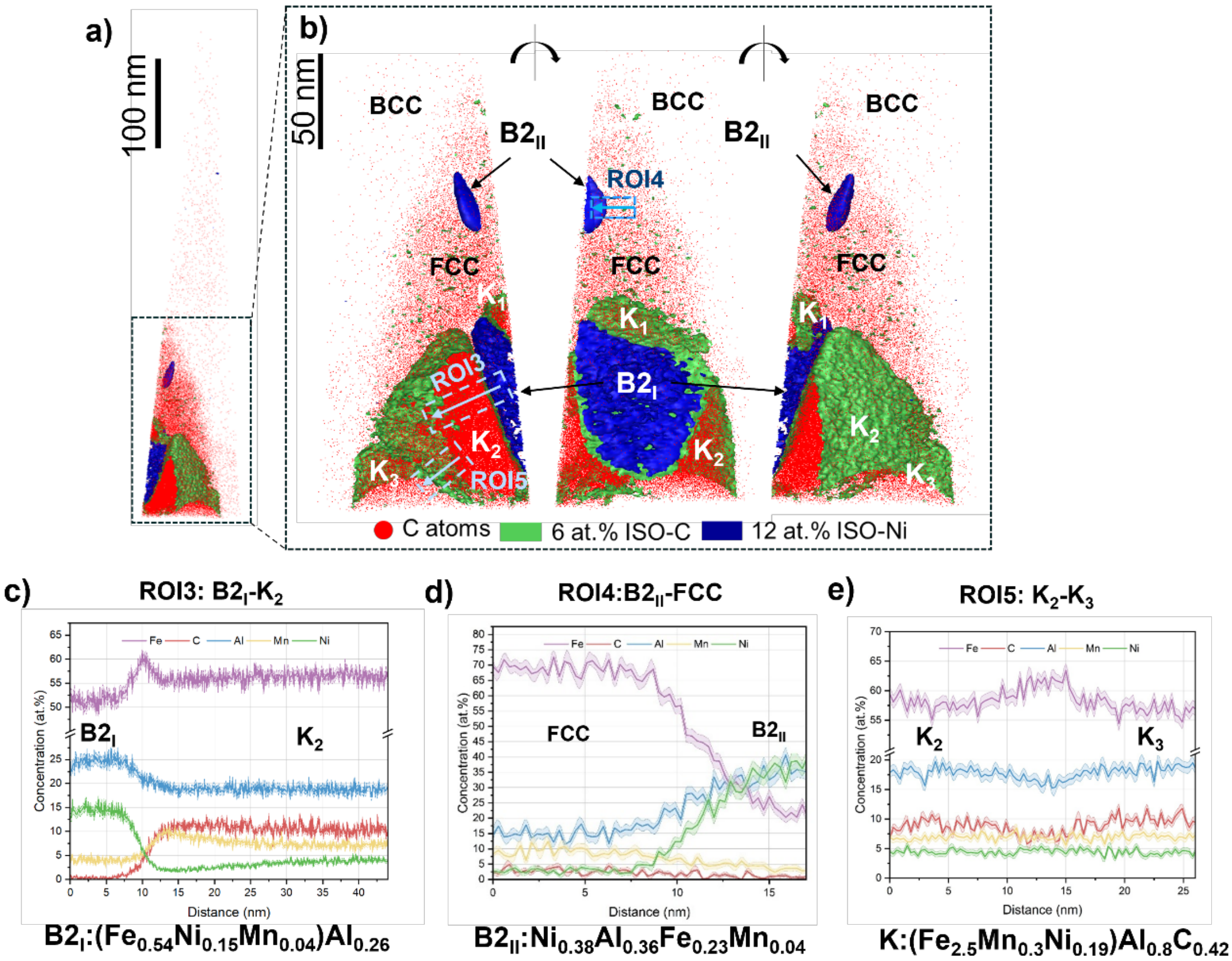

**Figure 3.** APT analysis of intermetallic $\kappa$ and B2 phases within the FCC matrix near the phase boundary a) 3D reconstruction of the intermetallic phases in APT maps of C atoms (red), b) magnified intermetallic phases/FCC matrix region with C iso-concentration surface as 6 at.% (green) and Ni iso-concentration surface as 12 at.% (blue) of the investigated Al-alloyed steel after the ageing at 800 °C for 15 min and the 1D concentration profiles associated with selected c) ROI 3 across $B2_I$-$K_2$; d) ROI 4 across $B2_{II}$-FCC; e) ROI 5 across $K_2$-$K_3$.

### 3.2 DFT calculation for the interaction between $\kappa$ and B2 phases

Based on APT analysis, the $\kappa$ phase in the studied alloy exhibits a non-stoichiometric composition, namely Fe-site substituted by minor amounts of Mn and Ni, as well as the presence of both Al and C vacancies. While Mn and Al are recognized as stabilizing elements in the $\kappa$ phase [52,53], the role of Ni in the $\kappa$ phase remains unclear and needs further investigation. To provide insight into the incorporation of Ni into the $\kappa$ phase, a hypothetical reaction scenario in the FCC matrix can be considered, where Ni from the matrix is involved in altering the composition of a pristine $\kappa$ phase. The following reaction would be the first to investigate:

$$Ni + Fe_3AlC \rightarrow Fe_\gamma + Fe_2NiAlC$$

where the subscript $\gamma$ denotes the $\gamma$ polymorph of Fe, i.e., FCC. The enthalpy change ($\Delta H$) of such a reaction amounts to –203.1 kJ/mol. The negative sign suggests that such a reaction is rather thermodynamically favorable. The main driving force for this reaction is the formation of FCC. Using materials-science language, this thermodynamically favorable reaction would be interpreted as the interactions of the *κ* phase and the FCC matrix. The *κ* phase would be inclined to exchange Fe for Ni from the Ni-containing FCC matrix. The energy of Ni on the left side would be replaced by the free energy of the Fe–Ni solid solution, but the relation should still hold as one does not expect any Ni in the pristine *κ* phase, and the formation of FCC (or Fe-rich FCC solid solution) is always favorable and driving the reaction.

More importantly, two types of B2 phase were detected by APT, and their compositions appear to depend on whether they form close to the *κ* phase or not. Therefore, it is also worth checking the interactions between *κ* and the B2 phases to understand how the positional relationship influences their co-precipitation behavior. The following reaction is considered:

$$NiAl + Fe_3AlC \rightarrow FeAl + Fe_2NiAlC$$

Both NiAl and FeAl are intermetallic phases crystallizing into the B2 structure. Such a reaction would be helpful to interpret the consequences of contacting B2 and *κ* phases. The $\Delta H$ of such a reaction reaches –49.11 kJ/mol, signifying another thermodynamically favorable process. In this vein, it would be inferred that the B2 phase contacting a *κ* phase would possess less Ni in the chemical composition, compared to the isolated B2 phase. Summarizing the results of the two reactions above, we would draw the conclusion that the incorporation of Ni into the *κ* phase is driven by rather strong thermodynamic driving forces.

As the microconstituents were formed after heat treatment at 1073 K, it would be meaningful to perform calculations at finite temperatures to reveal thermodynamic properties of the above reaction. To do so, phonon frequencies were calculated via a finite displacement scenario for $Fe_3AlC$, $Fe_2NiAlC$, NiAl, and FeAl, respectively. The results are plotted in **Figure 4**a.

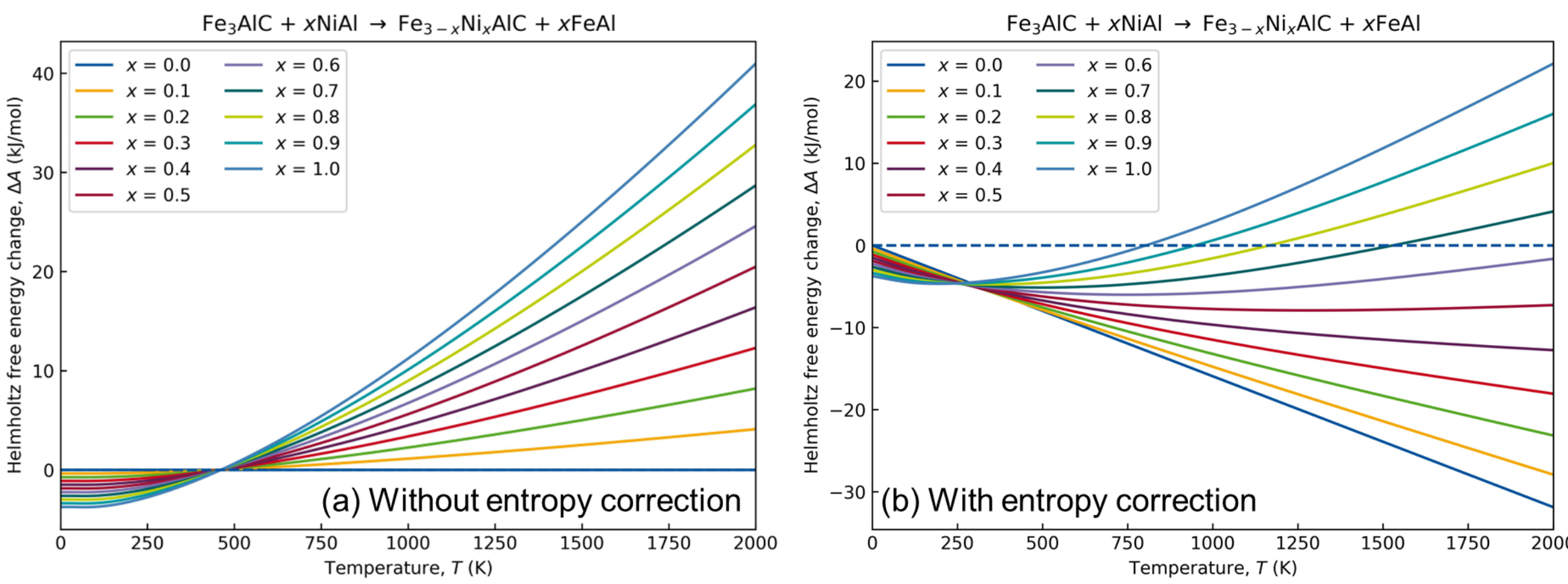

**Figure 4.** Temperature dependence of the Helmholtz free energy change ($\Delta A$) for the reaction: $Fe_3AlC + xNiAl \rightarrow Fe_{3-x}Ni_xAlC + xFeAl$ (x = 0–1) plotted (a) without and (b) with entropy corrections.

To elaborate on the solid solution, linear interpolations were performed to estimate the effects of smaller amounts of substitution in $Fe_{3-x}Ni_xAlC$. It can be found in Figure 4a that the reaction is not thermodynamically favorable when the temperature exceeds ~ 500 K. Smaller amounts of substitution would alleviate the unfavourability, but the crossover point for the course of Helmholtz free energy change remains unchanged. Nevertheless, it is worth noting that the phonon calculation on the level of harmonic approximation did not take the following three entropic contributions into consideration: the mixing entropy caused by forming the solid solution, the magnetic entropy resulting from the disordering of magnetic moments, and the electronic entropy stemming from electronic excitations at finite temperatures. Therefore, corresponding corrections on the entropy terms were also performed. The mixing entropy ($\Delta S_{mix}$) was estimated by:

$$\Delta S_{mix} = -R\Sigma c_i \ln(c_i)$$

where $R$ is the ideal gas constant and $c_i$ is the concentration of component $i$. Similarly, the magnetic entropy ($\Delta S_{mag}$) can be estimated by:

$$\Delta S_{mag} = R\ln(m + 1)$$

where $m$ is the total magnetic moment in the system. As regard the electronic entropy ($\Delta S_{ele}$), the magnitude of such effect is quantified by:

$$\Delta S_{ele} = k_B \int N(E)[f\ln f + (1-f)\ln(1-f)]dE$$

where $k_B$ is the Boltzmann constant, $N(E)$ is the density of states and $f$ stands for the Fermi-Dirac distribution expressed by:

$$f = 1 / \{1 + \exp[(E-E_F)/k_BT]\}$$

The results of the Helmholtz free energy corrected by manual intervention of the entropy term are shown in Figure 4b. Even though the free energy change ($\Delta A$) of $Fe_2NiAlC$ is still positive at 1073 K, the trend that a smaller amount of substitution of Fe for Ni would be favorable, and the formation of such a phase is favored by a more appropriate description of entropy.

After incorporating Ni into the pristine $\kappa$ phase, the changes in chemical bonds in the $\kappa$ phase are also valuable in understanding the stability of pristine and substituted $\kappa$ phases. For the pristine $\kappa$ phase, as shown in **Figure 5**, it can be found that the strongest stabilization effects arise from the Fe–C (six-fold) and Fe–Al (twelve-fold) interactions, according to the integrated COHP (ICOHP). Therefore, iron atoms are identified as critical to understand the stability of the $\kappa$ phase. After substitution on the Fe site, the stability of the system would be significantly perturbed. From Figure 5, it can be observed that, using the rigid band model, when the $E_F$ is moved downwards, the stability of the system improves, as there are mainly bonding/non-bonding levels underneath the $E_F$. This explains the great stability found in $Mn_3AlC$ and $(Fe,Mn)_3AlC$.

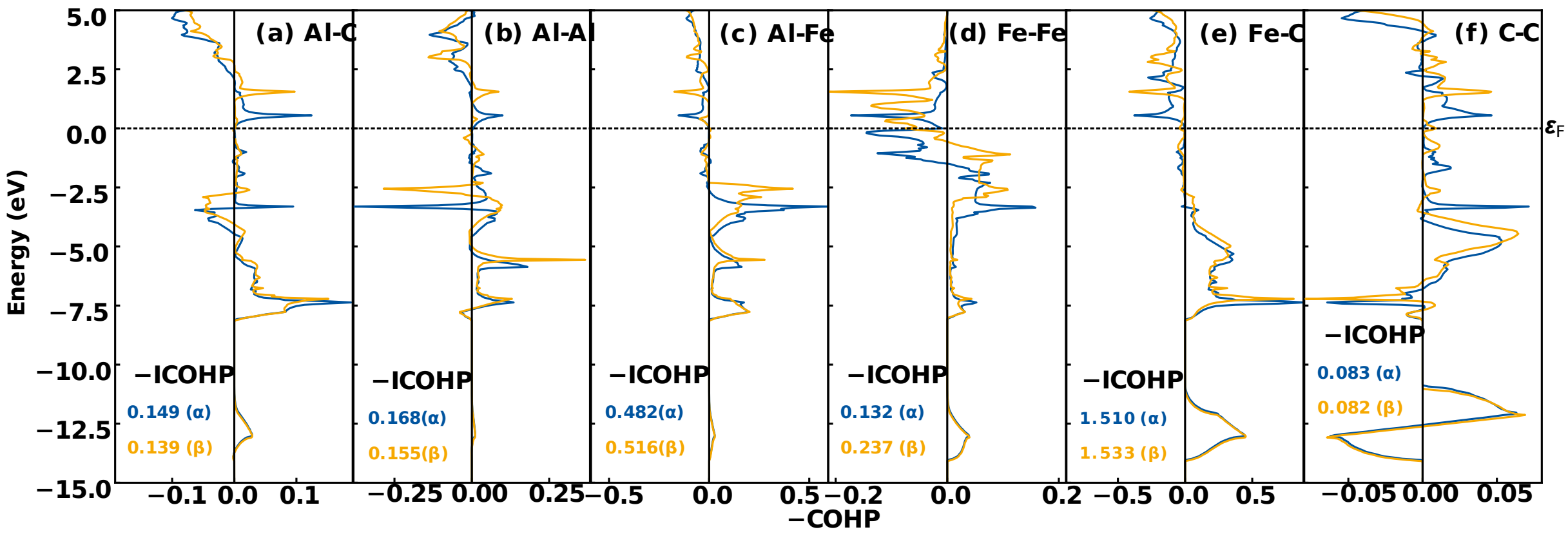

**Figure 5.** COHP courses of (a) Al–C, (b) Al–Al, (c) Al–Fe, (d) Fe–Fe, (e) Fe–C, and (f) C–C interactions in $Fe_3AlC$. Blue and yellow are used for the shortest interatomic distance, whereas green and red are used for the second shortest interatomic distance. Blue and green are used for spin majority channel ($\alpha$) and yellow and red are used for spin minority channel ($\beta$).

However, upon substituting Fe for Ni in $Fe_3AlC$, the Fermi level would be moved upwards. As a consequence, the unoccupied antibonding levels above the $E_F$ would be populated, especially in cases of Fe–C and Fe–Al interactions. It is intuitive to guess that Ni would cause detrimental effects on the stability of $\kappa$ phase. The COHP results for $Fe_2NiAlC$ are shown in **Figure 6**. The introduction of Ni to replace Fe did cause a significant drop in ICOHP. The ICOHP values for Ni–C interactions and Ni–Al interactions are 2.404 and 0.757 eV/bond, respectively, circa 20% smaller than the values for Fe–C and Fe–Al in $Fe_3AlC$. The populated antibonding levels can be observed in Ni–C interactions and Ni–Al interactions, leading to the decrease in ICOHP values compared to the Fe counterparts. Interestingly, incorporation of Ni into $Fe_3AlC$ results in larger ICOHP values for Fe–C and Fe–Al interactions in $Fe_2NiAlC$, counteracting the detrimental effects caused by Ni. Therefore, we would draw the conclusion that introducing Ni into the crystal structure of the $\kappa$ phase causes a decrease in stability, but its detrimental effects on the crystal structure are also alleviated by its presence. The net influence would be that this $\kappa$ phase presents a certain tolerance to the incorporation of Ni.

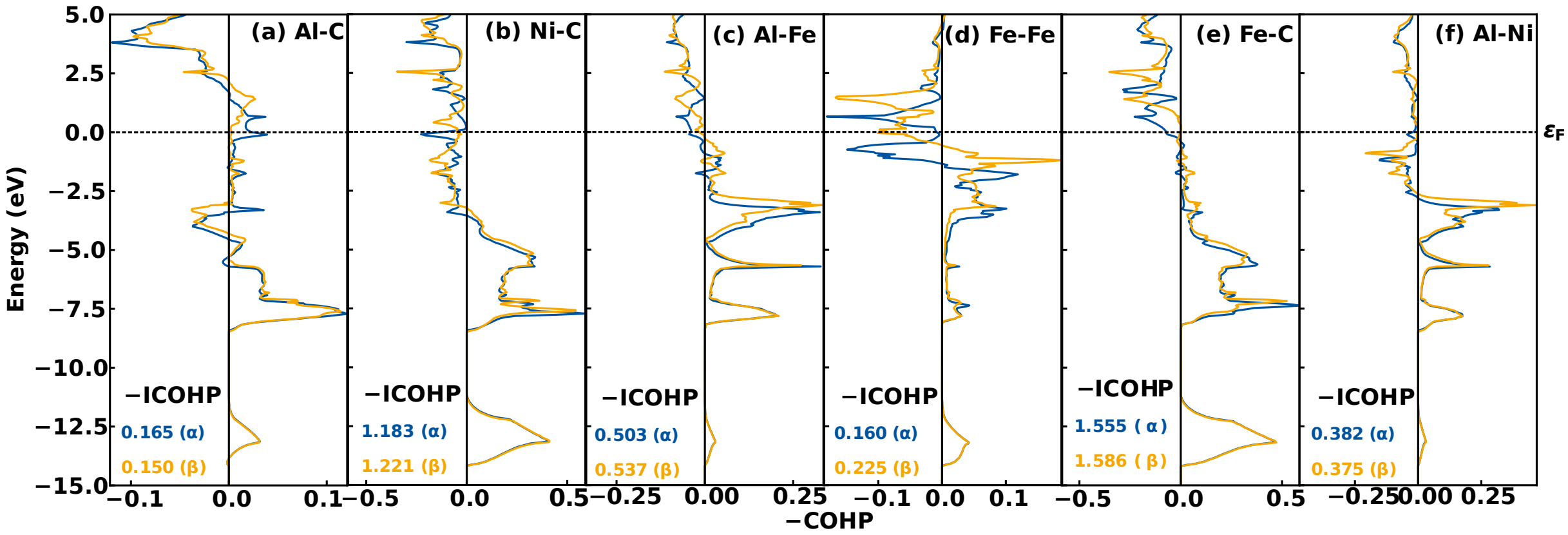

**Figure 6.** COHP courses of (a) Al–C, (b) Ni–C, (c) Al–Fe, (d) Fe–Fe, (e) Fe–C, and (f) Al–Ni interactions in $Fe_2NiAlC$. Blue and yellow are used for the shortest interatomic distance, whereas green and red are used for the second shortest interatomic distance. Blue and green are used for spin majority channel ($\alpha$) and yellow and red are used for spin minority channel ($\beta$).

In addition, the ionic contribution to the stability of the system is also non-negligible. As a measure of ionicity, the Madelung energy, derived from Löwdin charges, significantly lowers (i.e., greater stability) from –1409.64 kJ/mol (for pristine $\kappa$ phase) to –1576.56 kJ / mol (for Ni-substituted $\kappa$ phase). Such a significant lowering in Madelung energy suggests that the ionicity is providing stronger stabilizing effects to the Ni-substituted $Fe_3AlC$. The reason is that Ni atoms bear more positive charges than Fe on the equivalent positions.

In summary, the incorporation of Ni into the $\kappa$ phase exhibits both beneficial and detrimental effects. On the positive side, (i) it is driven by strong thermodynamical driving forces that favor Ni incorporation, (ii) it enhances the Fe–C and Fe–Al interactions in $Fe_2NiAlC$, thereby counteracting the weakening effect introduced by Ni, and (iii) it significantly lowers the Madelung energy, suggesting stronger stabilizing contributions from the ionic interactions due to the higher positive charge borne by Ni compared to Fe. On the negative side, the substitution of Fe by Ni leads to the occupation of antibonding levels and consequently decreases the overall stability of the $\kappa$ phase. Taking all these effects together, it is suggested that the κ phase exhibits a certain tolerance to Ni incorporation, where the stabilizing contributions partly alleviate the destabilizing influence.

### 3.3 Elemental redistribution between $\kappa$ and B2 phase during their co-precipitation

To further understand how $\kappa$ and B2 phases interact during their co-precipitation, this section focuses on the role of positional relationship on their chemical characteristics and elemental redistribution, based on combined insights from APT analysis and DFT calculations. Some studies on precipitates in steels confirm that the spatial position of precipitates during co-precipitation significantly changes their chemical features and elemental distribution. For instance, Danoix et al. [54] demonstrated that intermetallic NiAl-type precipitates directly influence nucleation and distribution of secondary hardening carbides in medium-carbon martensitic steels, changing nucleation sites from dislocations to intermetallic precipitates and consequently altering their chemical distribution. Similarly, in multicomponent Fe–Cu steels, the co-precipitation of Cu-rich precipitates with carbides (NbC or $Fe_3C$) results in mutual elemental redistribution at their heterophase interfaces. This interfacial redistribution alters the local chemistry and segregation behavior compared to isolated precipitates, leading to the formation of two distinct populations of NiAl-type phases, one containing Mn and one essentially Mn-free, at the Cu-rich precipitate interfaces [55]. In the present study, APT results reveal that when $\kappa$ and B2 precipitates are close to each other (e.g., like $B2_I$ and $K_2$), the $B2_I$ exhibits different chemical characteristics compared to the isolated $B2_{II}$. Since Al is known to stabilize both $\kappa$ and B2 phases, it is necessary to figure out whether there is competition for Al between $\kappa$ and B2 phases during their co-precipitation.

According to the APT results, the average Al content in $\kappa$ precipitates within the interior of the FCC matrix is 18.42 ± 0.23 at.%, while the $\kappa$ precipitate ($K_2$ in Figure 3) forming together with $B2_I$, shows a similar Al content in an average of 18.06 ± 0.91 at.%, and another $\kappa$ precipitate $K_3$ exhibits an Al content of 18.56 ± 0.93 at.%. The comparison of $\kappa$ precipitates located in different regions shows negligible differences in Al content. This indicates that $\kappa$ maintains a stable Al content regardless of whether it is near a B2 phase or not. However, $B2_I$, which co-precipitates closely adjacent to $\kappa$ precipitates, contains much less Al and Ni than the isolated $B2_{II}$. This contrast indicates that during co-precipitation, the $\kappa$ phase may have a stronger ability to obtain Al from the FCC matrix than the B2 phase does. Further supporting this inference, the FCC matrix surrounding co-precipitated $\kappa$ and B2 phases (near the FCC/BCC phase boundary) obviously shows decreased Al and Ni contents, around 2 at.% lower, compared to the matrix located in the interior region where only $\kappa$ precipitates are present. These observations pinpoint that both Al and Ni in $B2_I$ appear to be primarily supplied from the surrounding FCC matrix. However, because the $\kappa$ phase tends to retain its Al content even in proximity to the B2 phase, it exhibits a higher capability of attracting Al from the FCC matrix, resulting in less Al within $B2_I$.

The lower Ni content being characteristic for $B2_I$ can also be explained by its interaction with the adjacent $\kappa$ phase. Based on DFT results, the $\kappa$ phase can tolerate a limited amount of Ni substitution on Fe sites without compromising stability. The isolated $B2_{II}$, which nucleates and grows away from any $\kappa$ precipitates, shows a nearly 1:1 Ni:Al ratio, matching the composition of a typical NiAl-type B2 phase. As it continues to grow, it gradually absorbs Fe from the surrounding FCC matrix, eventually forming a NiAl-type B2 containing Fe. In contrast, $B2_I$, which forms in direct proximity to $\kappa$, has lower Al, Ni content and elevated Fe content. This deviation from the ideal NiAl composition is likely due to local solute redistribution driven by $\kappa$–B2 interactions. The formation of $\kappa$ leads to Al depletion in the surrounding FCC matrix, reducing the Al incorporation into the further growth of NiAl-type B2. Moreover, the Fe content in $\kappa$ precipitates next to B2 is approximately 3.8 at.% lower than in $\kappa$ precipitates within the interior of the FCC matrix, indicating Fe redistribution between the $\kappa$ and B2 phases during their co-precipitation. The Ni content within the $\kappa$ phase further supports this scenario. $\kappa$ precipitates within the interior of the FCC grains having an average Ni content of 4.61 ± 0.19 at.%, slightly lower than that of the surrounding FCC matrix (5.34 ± 0.35 at.%). This makes sense because, without B2 precipitation nearby, the $\kappa$ phase can only incorporate Ni from the matrix and thus should contain less Ni than the matrix itself. Near co-precipitated B2 phases, however, $\kappa$ precipitates maintain a similar Ni content (4.35 ± 0.32 at.%), despite a significant drop of Ni content (3.2 ± 0.26 at.%) in the adjacent matrix. This observation indicates that during co-precipitation, these $\kappa$ precipitates obtain Ni from an additional source besides the local FCC matrix. In other words, the nearby B2 phase likely supplies extra Ni, allowing $\kappa$ precipitates to reach Ni levels even higher than the matrix. Taking into account the lower Ni content in $B2_I$, these results indicate that Fe–Ni atom exchange takes place between the *κ* and B2 phases when they co-precipitate in proximity.

This Fe–Ni atom exchange between *κ* and $B2_I$ aligns well with the DFT results, which exhibit a thermodynamically favorable state. At 800 °C, simulations show that *κ* and B2 phases in contact promote elemental redistribution, especially in terms of Fe and Ni. The *κ* phase tolerates limited Ni substitution on Fe sites without compromising its stability. At the same time, B2 can incorporate excess Fe and form a stable FeAl-type phase (B2`), even with lower Al content [49]. This leads to a change from a NiAl-type to a FeAl-type B2 phase, which is thermodynamically favorable in the outskirts of *κ*. The calculated reaction at 1073 K:

$$x\text{NiAl} + \text{Fe}_3\text{AlC} \rightarrow x\text{FeAl} + \text{Fe}_{3\text{-}x}\text{Ni}_x\text{AlC}\ (x = 0\text{–}0.8),\ \Delta A < 0\ \text{kJ/mol}$$

confirms that such Fe–Ni exchange is energetically feasible.

Based on these results, two distinct precipitation pathways can be outlined. (i) Isolated B2 precipitation (forms away from the *κ* phase): the B2 precipitates nucleating away from the *κ* phase grow through absorption of Ni and Al from the FCC matrix. During growth, they retain a NiAl-type composition, although a considerable amount of Fe is inevitably incorporated from the surrounding Fe-based FCC matrix, resulting in a Fe-containing NiAl-type B2 phase. (ii) Close co-precipitation with the *κ* phase: when the B2 precipitate forms together with the *κ* phase in the FCC matrix, competitive partitioning of Al from the matrix favors the *κ* phase, while Fe–Ni redistribution between *κ* and B2 occurs. This promotes the evolution of B2 towards a FeAl-type composition and simultaneously allows *κ* to accommodate Ni, consistent with DFT predictions of favorable Fe–Ni atom exchange.

In summary, both experimental and DFT results support that the coexistence of *κ* and B2 phases during precipitation can follow two distinct mechanisms depending on their relative positions. The isolated B2 precipitate evolves along the pathway: NiAl → Fe-containing NiAl. The B2 precipitates that form together with the *κ* precipitates, however, undergo cooperative solute redistribution with the *κ* phase, involving Al competition and Fe–Ni atom exchange. These results highlight that the interaction between *κ* and B2 phases during co-precipitation in the FCC matrix of Fe-Mn-Al-Ni-C lightweight steel is not coincidental but reflects two different precipitation mechanisms governed by their positional relationship, involving elemental redistribution and accommodation between these intermetallic phases.

## 4. Conclusions

In the present work, nanoscale characterization, near-atomic investigation, and DFT calculations were combined to systematically investigate the co-precipitation behavior of the *κ* and B2 phases within the FCC matrix of the investigated Al-alloyed lightweight steel (Fe–10Al–7Mn–6Ni–1C, wt.%) aged at 800 °C for 15 minutes. The study specifically evaluated how the positional close co-precipitation of *κ* and B2 phases affects their chemical characteristics and phase stability.

Experimentally, the *κ* precipitates within the FCC matrix were identified with a stoichiometry of $(Fe_{2.7}Mn_{0.27}Ni_{0.2})Al_{0.8}C_{0.4}$ in an L’12-structure, while the *κ* precipitates near the FCC/BCC phase boundary showed a similar stoichiometry of $(Fe_{2.5}Mn_{0.3}Ni_{0.19})Al_{0.8}C_{0.42}$. These *κ* precipitates exhibited minor substitutions at Fe sites by Mn and Ni, along with notable Al and C vacancies. Additionally, two distinct types of B2 precipitates were observed: $B2_I$, forming close to *κ* precipitates, was identified as a FeAl-type precipitate; Isolated $B2_{II}$, which can be characterized as a NiAl-type B2 precipitate. The stable Al content (≈18.4 at.%) in *κ* precipitates, regardless of their positional proximity to B2 precipitates, indicated a competitive advantage for Al acquisition by *κ* over B2. Notably, *κ* precipitates next to B2 precipitates contained approximately 3.8 at.% less Fe compared to *κ* precipitates without neighboring B2 precipitates, revealing clear evidence of Fe redistribution between the co-precipitating *κ* and B2 phases.

Simulation results from DFT calculations support these observations, showing that the *κ* phase can accommodate a small amount of Ni substitution at Fe sites without losing structural stability, in agreement with the experimental evidence of Ni incorporation. In addition, the calculations indicate that Fe–Ni atom exchange between *κ* and B2 phases is thermodynamically favorable at 800 °C, rationalizing the compositional differences between the co-precipitated and the isolated B2 precipitate within the FCC matrix.

These findings indicate that although *κ* and B2 phases compete for Al from the FCC matrix, their interaction during co-precipitation is essentially cooperative by mutual redistribution of Fe and Ni. The B2 phases in proximity to *κ* phases can incorporate higher Fe levels under Al-depleted conditions, shifting their composition from a NiAl-type towards a FeAl-type structure. Therefore, positional proximity during co-precipitation of *κ* and FeAl-type B2 phases significantly influences their chemical characteristics and may facilitate their mutual stabilization. Ultimately, these insights provide a fundamental understanding of *κ*–B2 interactions during co-nanoprecipitation and offer guidance for developing optimized co-precipitation strategies to tailor the microstructure and mechanical properties of Al-alloyed Fe-Mn-Al-Ni-C lightweight steels.

**Acknowledgements**

B.Z. and Y.W. contributed equally to this work. Financial support from the Deutsche Forschungsgemeinschaft (DFG) under grant no. 314595867 and 490856143 is gratefully acknowledged. T.N. acknowledges DFG funding under grant no. 495468531. The author B.Z. would also like to express his sincere appreciation to the Deutscher Akademischer Austauschdienst (DAAD) for supporting his exchange at the Indian Institute of Technology Madras (IITM), and to Prof. K. G. Pradeep, Department of Metallurgical and Materials Engineering, IITM, for his kind assistance in performing the TKD and TEM characterizations.

**Conflict of Interest**

The authors declare that they have no conflict of interest.

**Data Availability Statement**

The data that support the findings of this study are available from the corresponding author upon reasonable request.

**Near-atomic investigation on the elemental redistribution during co-precipitation of nano-sized kappa phase and B2 phase in an Al-alloyed lightweight steel**

ToC figure

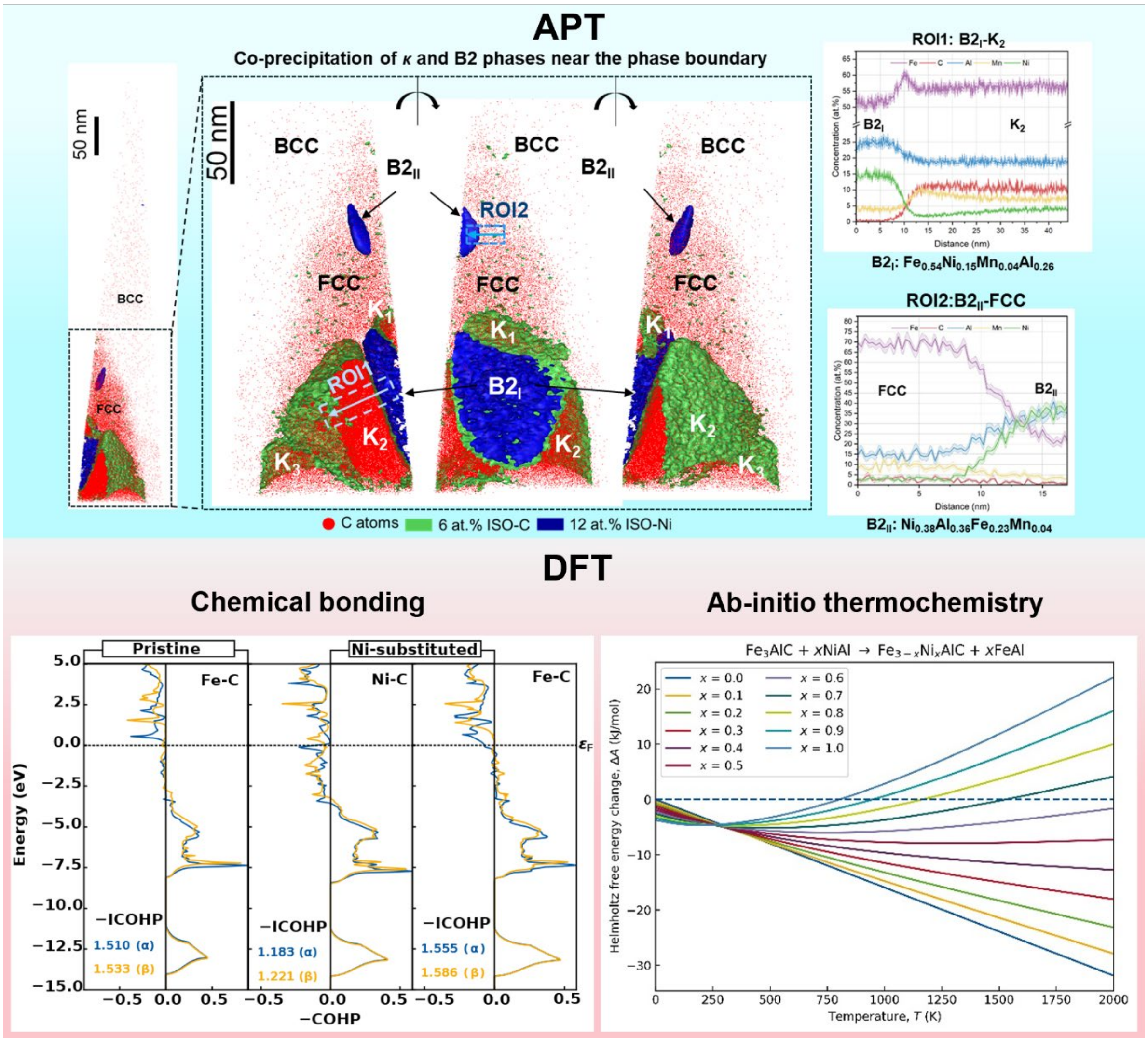

Supporting information

# Near-atomic investigation on the elemental redistribution during co-precipitation of nano-sized kappa phase and B2 phase in an Al-alloyed lightweight steel

*Bowen Zou[a], Yixu Wang[a,b], Xiao Shen[a], Philipp Krooß[c], Thomas Niendorf[c], Richard Dronskowski[b], Wen-wen Song[a*]*

[a]: Department of Granularity of Structural Information in Materials Engineering, Institute of Materials Engineering, University of Kassel, Mönchebergstr. 3, 34125 Kassel, Germany

[b]: Chair of Solid-State and Quantum Chemistry, Institute of Inorganic Chemistry, RWTH Aachen University, Landoltweg 1, 52056 Aachen, Germany

[c]: Department of Metallic Materials, Institute of Materials Engineering, University of Kassel, Mönchebergstr. 3, 34125 Kassel, Germany

[*] Corresponding author: song@uni-kassel.de

**Table S1.** Atomic chemistry of precipitates/matrix (P/M) and corresponding stoichiometry in the FCC matrix interior in the investigated Al-alloyed steel upon annealing at 800 °C for 15 minutes followed by water quenching ($K_A$: Average concentration of *κ* phase, $FCC_A$: Average concentration of FCC channel).

| *κ* phase precipitation in the FCC matrix interior | | | | | | |
|---|---|---|---|---|---|---|
| P/M | Stoichiometry | Element concentration in at.% (±S.E.) | | | | |
| | | Fe | C | Mn | Al | Ni |
| $K_1$ | $(Fe_{2.7}Mn_{0.28}Ni_{0.18})Al_{0.8}C_{0.42}$ | 62.20 ±1.02 | 9.29 ±0.60 | 6.32 ±0.50 | 17.77 ±0.85 | 4.09 ±0.41 |
| $K_2$ | $(Fe_{2.7}Mn_{0.26}Ni_{0.21})Al_{0.8}C_{0.40}$ | 61.68 ±1.56 | 9.12 ±0.91 | 5.97 ±0.75 | 18.11 ±1.31 | 4.83 ±0.68 |
| $K_3$ | $(Fe_{2.7}Mn_{0.27}Ni_{0.22})Al_{0.8}C_{0.39}$ | 61.26 ±1.46 | 8.85 ±0.83 | 6.18 ±0.71 | 18.36 ±1.22 | 4.95 ±0.64 |
| $K_4$ | $(Fe_{2.7}Mn_{0.28}Ni_{0.21})Al_{0.8}C_{0.39}$ | 61.56 ±1.64 | 8.93 ±0.95 | 6.27 ±0.81 | 18.09 ±1.38 | 4.71 ±0.70 |
| $K_5$ | $(Fe_{2.6}Mn_{0.26}Ni_{0.19})Al_{0.8}C_{0.41}$ | 60.75 ±0.69 | 9.61 ±0.42 | 6.11 ±0.34 | 18.56 ±0.59 | 4.37 ±0.29 |
| $K_6$ | $(Fe_{2.6}Mn_{0.26}Ni_{0.20})Al_{0.8}C_{0.39}$ | 60.67 ±0.71 | 9.17 ±0.41 | 6.10 ±0.34 | 18.87 ±0.60 | 4.64 ±0.30 |
| $K_7$ | $(Fe_{2.6}Mn_{0.25}Ni_{0.20})Al_{0.8}C_{0.40}$ | 60.41 ±0.71 | 9.49 ±0.42 | 6.08 ±0.34 | 19.15 ±0.59 | 4.64 ±0.29 |
| $K_A$ | $(Fe_{2.7}Mn_{0.27}Ni_{0.20})Al_{0.8}C_{0.40}$ | 61.16 ±0.45 | 9.21 ±0.26 | 6.15 ±0.23 | 18.42 ±0.23 | 4.61 ±0.19 |
| $FCC_1$ | - | 69.32 ±1.52 | 3.07 ±0.55 | 6.32 ±0.78 | 15.29 ±1.26 | 5.45 ±0.73 |
| $FCC_2$ | - | 68.28 ±0.72 | 3.40 ±0.27 | 6.38 ±0.37 | 15.71 ±0.59 | 5.55 ±0.35 |
| $FCC_3$ | - | 69.19 ±1.48 | 3.32 ±0.53 | 6.45 ±0.77 | 15.64 ±1.23 | 5.02 ±0.68 |
| $FCC_A$ | - | 68.93 ±0.75 | 3.26 ±0.27 | 6.38 ±0.39 | 15.55 ±0.62 | 5.34 ±0.35 |

**Table S2.** Atomic chemistry of precipitates/matrix (P/M) and corresponding stoichiometry near FCC/BCC phase boundary in the investigated steel upon annealing at 800 °C for 15 minutes, followed by water quenching ($K_A$: Average concentration of *κ* phase).

| Co-precipitation of intermetallic phases in FCC matrix near the FCC/BCC phase boundary | | | | | | |
|---|---|---|---|---|---|---|
| P/M | Stoichiometry | Element concentration in at.% (±S.E.) | | | | |
| | | Fe | C | Mn | Al | Ni |
| $K_2$ | $(Fe_{2.5}Mn_{0.3}Ni_{0.20})Al_{0.8}C_{0.40}$ | 57.36 ±1.10 | 9.12 ±0.64 | 6.84 ±0.56 | 18.06 ±0.91 | 4.41 ±0.45 |
| $K_3$ | $(Fe_{2.4}Mn_{0.3}Ni_{0.18})Al_{0.8}C_{0.43}$ | 56.79 ±1.12 | 9.95 ±0.67 | 7.06 ±0.58 | 18.56 ±0.93 | 4.29 ±0.46 |
| $K_A$ | $(Fe_{2.5}Mn_{0.3}Ni_{0.19})Al_{0.8}C_{0.42}$ | 57.28 ±0.78 | 9.54 ±0.46 | 6.95 ±0.40 | 18.31 ±0.65 | 4.35 ±0.32 |
| $B2_I$ | $Fe_{0.54}Ni_{0.15}Mn_{0.04}Al_{0.26}$ | 51.46 ±1.19 | 0.35 ±0.13 | 4.01 ±0.49 | 24.79 ±1.08 | 14.54 ±0.84 |
| $B2_{II}$ | $Ni_{0.38}Al_{0.36}Fe_{0.23}Mn_{0.04}$ | 22.22 ±1.94 | 0.65 ±0.38 | 3.66 ±1.13 | 34.78 ±2.83 | 37.25 ±2.80 |
| FCC | - | 69.09 ±0.67 | 3.17 ±0.26 | 9.34 ±0.43 | 13.62 ±0.55 | 3.20 ±0.26 |
| BCC | - | 71.89 ±0.67 | 0.04 ±0.02 | 5.40 ±0.32 | 17.47 ±0.59 | 4.49 ±0.29 |